\documentstyle[12pt]{article}

\begin{document}
\title{Multiple Electromagnetic Excitation in Fast Peripheral
	 Heavy Ions Collisions}
\author{\underline{S. Typel}\thanks{E-Mail: stypel@ulb.ac.be} \\
	  Physique Nucl\'{e}aire Th\'{e}orique
	  et Physique Math\'{e}matique, \\
	  Universit\'{e} Libre de Bruxelles,
	    Campus de la Plaine CP229, \\
	  Bvd.~du Triomphe, B-1050 Bruxelles, Belgium \\
	  G. Baur \\
	  Institut f\"{u}r Kernphysik,
	    Forschungszentrum J\"{u}lich GmbH, \\
	  D-52425 J\"{u}lich, Germany}
\date{}
\maketitle
\clearpage
\begin{abstract}
We study the corrections of first order electromagnetic excitation due
to higher order electromagnetic interactions. An effective operator is
introduced which takes these effects into account in the sudden
approximation. Evaluating the matrix-elements of this operator between
the relevant states corrections to the first order result are obtained
in a simple way. As an example we discuss the excitation of the first
excited state in ${}^{11}$Be. It tends to improve the
agreement between experiment and theory.
\end{abstract}
\clearpage
Electromagnetic Excitation in the energy domain of several
tens of MeV/u up to relativistic energies is a growing field of study.
The cross-section can become large and irreducible nuclear effects
can be kept under control. With increasing beam energy the equivalent
photon spectrum becomes harder, and also particle-unstable states
can be reached. The Coulomb dissociation
${}^{11}$Li $\to$ ${}^{9}$Li + 2n and ${}^{14}$O $\to$ ${}^{13}$N + p,
which is also astrophysically relevant, are examples \cite{Bau94}.
Recently, bound states were also excited and their (Doppler shifted)
de-excitation $\gamma$-rays were measured. A large deformation of the
neutron-rich nucleus ${}^{32}$Mg was recently deduced from a measurement
of the $2^{+} \to 0^{+}$ 885~keV transition to the ground state
after medium energy electromagnetic excitation \cite{Mot95a}.
The 320~keV $\frac{1}{2}^{-} \to \frac{1}{2}^{+}$ $\gamma$-transition
in ${}^{11}$Be was recently observed. The measured cross-section
for the ${}^{11}$Be ($\frac{1}{2}^{+} \to \frac{1}{2}^{-}$) Coulomb
excitation was found to be noticeably less than expected
from the known lifetime and $1^{st}$ order pure Coulomb excitation
\cite{Ann94}.
Apart from possible nuclear and Coulomb-nuclear interference effects,
a possible reason for this discrepancy is the influence of higher order
electromagnetic interaction. It is the purpose of this letter
to describe a framework suitable for fast projectiles. E.g. the rather
loosely bound ${}^{11}$Be in its $1^{st}$ excited $\frac{1}{2}^{-}$ state
could easily be excited electromagnetically into the
continuum in a second
step \cite{Han94}.
\par
Electromagnetic excitation is mainly characterized by two parameters,
the adiabaticity parameter
\begin{equation}
 \xi = \frac{\omega b}{\gamma v}
\end{equation}
and the strength parameter
\begin{equation} \label{chidef}
 \chi_{fi}^{(E\lambda)} =
	\frac{Ze \langle f | {\cal M} (E \lambda \mu) | i \rangle}{
		  \hbar v b^{\lambda}}.
\end{equation}
The excitation energy is given by $\hbar \omega$, the impact parameter
in a straight-line approximation is denoted by $b$, and $\gamma=
\sqrt{1-(v/c)^{2}}$, where $v$ is the projectile velocity.
The target charge number is denoted by $Z$, and ${\cal M}(E \lambda \mu)$
the electric multipole operator.
\par
In Coulomb excitation below the barrier, multiple electromagnetic excitation
is usually treated in a coupled channels approach using the
relevant states from appropriate nuclear models,
like the harmonic vibrator or rigid rotor.
For a review see Ref.~\cite{Ald75}.
The situation for Coulomb excitation above the barrier becomes simpler,
because the excitations tend to be sudden.
While $\hbar \omega$ is restricted
to a few MeV, the adiabaticity parameter $\xi$ is typically less than 1,
for the important range of impact parameters $b > R_{1} + R_{2}$,
where $R_{1}$ and $R_{2}$ are the nuclear radii of projectile and
target. Thus fast collisions become the domain of the sudden
approximation \cite{Ald75}, or of a recently developed low-$\xi$
approximation \cite{Typ94a,Typ94b}. In this case it can be advantageous
to construct operators which take into account the influence of intermediate
states.
\par
For simplicity, let us use the straight-line approximation and the
dipole approximation. The first order excitation amplitude is given by
\cite{Typ94a}
\begin{equation}
 a_{fi}^{(1)}(E1,\xi)= -i \vec{q}(\xi) \cdot \langle f | \vec{r} |
			 i \rangle
\end{equation}
with
\begin{equation}
 \vec{q}(\xi) = \frac{2 Z Z_{eff}^{(1)} e^{2}}{\hbar v b}
		    \left( \begin{array}{c}
				\xi K_{1}(\xi) \\ 0 \\ i \gamma \xi K_{0}(\xi)
			     \end{array}  \right)
\end{equation}
where the projectile moves in the $z$-direction and
the impact parameter points to the $x$-direction.
The dipole effective charge is given by
\begin{equation}
  Z_{eff}^{(1)} = \frac{Z_{b}m_{c}-Z_{c}m_{b}}{m_{b}+m_{c}}
\end{equation}
using a model of two pointlike inert clusters $b$ and $c$ with
charge numbers $Z_{b}$, $Z_{c}$ and masses $m_{b}$ and $m_{c}$.
$K_{0}$ and $K_{1}$ are the modified Bessel functions.
For $\xi \to 0$ we get the classical Coulomb push
\begin{equation}
  \vec{q}(0) = \frac{2 Z Z_{eff}^{(1)} e^{2}}{\hbar v b} \vec{e}_{x}.
\end{equation}
In the limit $\xi << 1$, the
excitation amplitude can be evaluated easily to all orders in the
sudden approximation. It is given by
\begin{equation}
 a_{fi}^{sudden}(E1) = \langle f |
	  e^{-i \vec{q}(0) \cdot \vec{r}} | i \rangle .
\end{equation}
This could be generalized to higher multipolarities and trajectories
corrected for Coulomb deflection.
By comparing $a_{fi}^{(1)}(E1,\xi=0)$ and $a_{fi}^{sudden}(E1)$
the influence
of multiple electromagnetic excitation can be assessed. This is
a remarkably simple procedure, all intermediate states are included.
The reduction of the excitation probability due to higher order effects
is given by
\begin{equation}
 r = \left| \frac{a_{fi}^{sudden}(E1)}{a_{fi}^{(1)}(E1,\xi=0)} \right|^{2} .
\end{equation}
\par
For loosely bound states, e.g. in ${}^{11}$Be = ${}^{10}$Be + n,
we use simple model wave functions to reveal the characteristic
parameters. We choose two models, with the correct asymptotic behaviour.
Their differences give a feeling about the model dependence.
For the wave functions
we make the ansatz for the radial part in the initial state
(with orbital angular momentum $l_{i}=0$)
\begin{equation}
 u_{i}(r) = \sqrt{2 \alpha_{i}} \: \frac{\exp(-\alpha_{i} r)}{r}
\end{equation}
which corresponds to the solution of the Schr\"{o}dinger equation
with a $\delta$-like potential.
For the final state ($l_{f}=1$) we choose
\begin{equation}
 u_{f}^{I}(r) = \sqrt{2 \alpha_{f}} \: \frac{\exp(-\alpha_{f} r) }{r}
\end{equation}
and the more extended wave function
\begin{equation}
 u_{f}^{II}(r) = 2 \sqrt{\alpha_{f}^{3}} \:
 \exp(-\alpha_{f} r) ,
\end{equation}
respectively.
The constants $\alpha_{z}$ ($z=i,f$) are calculated from the
binding energies $E_{z}= \frac{\hbar^{2} \alpha_{z}^{2}}{2 \mu}$.
With these wave functions the calculated mean lifetimes of the
$\frac{1}{2}^{-}$ state are
110~fs and 97~fs, respectively. They are smaller than the experimental
value of ($166\pm 15$)~fs \cite{Mil83}. Despite this difference,
the ratio of higher order to first order effects can be given
with some confidence, where, e.g., common spectroscopic factors cancel out.
The absolute value depends on more
sophisticated details of the nuclear model (see, e.g., Ref.~\cite{Mil83}).
The reduction factor $r$ can be calculated
analytically. We get
\begin{equation}
 r^{I}=\left(\frac{3}{z^{2}}
   \left(1 - \frac{1}{z} \arctan z \right) \right)^{2}
\end{equation}
and
\begin{equation}
  r^{II}=\left(\frac{3}{2z^{3}}
   \left(\arctan z - \frac{z}{1+z^{2}} \right) \right)^{2} ,
\end{equation}
resp.,
where we have introduced the parameter
\begin{equation} \label{zdef}
 z=\frac{q(0)}{\alpha_{i}+\alpha_{f}}.
\end{equation}
The parameter $z$ is directly related to $\chi^{(E1)}_{fi}$
(Eq.~\ref{chidef}).
It describes the ratio of the strength of the Coulomb push $q$
and the ``looseliness'' of the system, and is a measure of the
importance of higher order effects.
The quantity $r$ is plotted in Fig.~1 for the two model wave-functions
described above.
\par
For large $b$ the sudden approximation fails ($\xi \to \infty$),
on the other hand, higher order effects diminish due to the
decrease of the strength parameter $\chi^{(E1)}_{fi}$.
The product
\begin{equation}
  \chi^{(E1)}_{fi} \cdot \xi =
	\frac{Ze \langle f | {\cal M} (E 1 \mu) | i \rangle \omega}{
		  \hbar \gamma v^{2}}
\end{equation}
is a very small number for low excitation energies $\hbar \omega$
and high projectile velocities $v$. The ranges of validity
for the first order calculation ($\chi$ small, $\xi$ arbitrary) and
the sudden approximation ($\xi$ small, $\chi$ arbitrary) overlap.
In a convenient and accurate
interpolation procedure we calculate the total cross section in the
following way
\begin{equation}
 \sigma^{(\infty)}
     = 2 \pi \int_{b_{min}}^{\infty} r(b) \:
     | a_{fi}^{(1)}(E1,\xi(b)) |^{2} \: b \: db
\end{equation}
which should be an accurate expression for all values of b in the
integrand.
For small impact parameters we have
$\xi \approx 0$ and the first order approximation cancels out in the
calculation of $\sigma^{(\infty)}$.
We compare it with the total cross section
in the first order calculation
\begin{equation}
 \sigma^{(1)} = 2 \pi \int_{b_{min}}^{\infty}
     | a_{fi}^{(1)}(E1,\xi(b)) |^{2} \: b \: db  .
\end{equation}
Both cross sections depend on the value of the minimum impact
parameter $b_{min}$. The change in the two cross sections will
be similar so that their ratio is less affected by a change in $b_{min}$.
If the sudden approximation is not well enough fulfilled,
one could use the low-$\xi$ approximation \cite{Typ94a,Typ94b}
which takes second order electromagnetic effects into account.
\par
For the ${}^{11}$Be Coulomb excitation
a reduction of the cross section from $(490\pm 50)$~mb
(expected from the first order calculation)
to the measured $(191\pm 26)$~mb was recently found
in an experiment at GANIL \cite{Ann94}.
Indeed, the parameter $z$ can become substantial in this case, and
higher order effects are not negligible. For a collision with
$b = 15$~fm and an energy of $45 \cdot A$~MeV ($v/c\approx 0.32$)
we have $z \approx 0.38$ and from Fig.~1
we see a substantial reduction of the excitation probability.
In a recent experiment at RIKEN \cite{Mot95b}, which is currently evaluated,
an energy of about $65 \cdot A$~MeV ($v/c \approx 0.38$)
for the ${}^{11}$Be was used.
This leads to a value of $z \approx 0.32$ for the same impact parameter
corresponding to a smaller reduction.
\par
The product $\chi^{(E1)}_{fi} \cdot \xi$ takes on the small
values $0.00375 (J_{i}M_{i}1 \mu | J_{f}M_{f})$ and
$0.00399 (J_{i}M_{i} 1 \mu | J_{f}M_{f})$, respectively, for
the two models and the GANIL conditions. At
$b= 10$~fm we get an excitation probability of 3.4\% and 3.9\%
in the first order calculation, decreasing with $b^{-2}$.
For impact parameters larger than $\frac{\gamma v}{\omega}
\approx 200$~fm the excitation becomes adiabatic and the
excitation probability drops off exponentially. The
excitation probability is small compared to 1 and the
non-conservation of unitarity in the first order
approximation will not affect the calculation of the
cross section.
\par
The apparent reduction of the B(E1)-value (due to higher order effects)
is given by
\begin{equation}
 R = \frac{\sigma^{(\infty)}}{\sigma^{(1)}}
\end{equation}
which is plotted in Fig.~2 as a function of the projectile
velocity $v$ for the two model wave functions
described above. 
We assume a minimum impact parameter
$b_{min}=10$~fm corresponding to a grazing collision of the
projectile and target. This will give an estimate of the
largest possible effect to be expected from the higher order
contributions.
We obtain a reduction, depending on the particular model chosen,
of $5.5 \%$ or $10.1 \%$ for the GANIL energy and $3.7 \%$ or $6.9 \%$
for the RIKEN energy.
The range of the reduction R in
the two models gives a
feeling of the reliabiliy of the results.
The use of more realistic wavefunctions is outside the scope
of the present work. The reduction of
the excitation probability for the $\frac{1}{2}^{-}$ state
by higher order effects will be accompanied by an
increase of the cross section in the breakup channel.
\par
In the dipole approximation, the lowest order correction was of
$3^{rd}$ order. E1-E2 excitation contributes already in second order.
Nevertheless, it is smaller than the third order E1-E1-E1 correction.
For low $\xi$ we can estimate the ratio of the two amplitudes as
\begin{equation}
 \frac{\mbox{E1-E2 amplitude}}{\mbox{E1-E1-E1 amplitude}} \approx
    \frac{1}{Z \alpha} \frac{v}{c} \frac{Z_{eff}^{(2)}}{(Z_{eff}^{(1)})^{2}}
    \approx 0.13
\end{equation}
for $Z=82$, $Z_{eff}^{(2)}=4/121$, and $Z_{eff}^{(1)}=4/11$ at the GANIL
energy. Thus the dipole approximation is reasonable, at least for a first
exploration.
\par
Possible higher order effects in the ${}^{32}$Mg intermediate energy
Coulomb Excitation \cite{Mot95a} can also be estimated.
Assuming a rigid rotor model, high energy Coulomb excitation was calculated
in the sudden approximation in Ref.~\cite{Bau89}.
The characteristic strength parameter is
\begin{equation}
 C = \frac{Ze^{2}}{\hbar v} \frac{Q_{0}}{2b^{2}}.
\end{equation}
Using the B(E2)-value found in Ref.~\cite{Mot95a} we have
(see eq.~6 of Ref.~\cite{Bau89}, where a factor $\pi$ is missing
on the rhs.)
\begin{equation}
 Q_{0}^{2} = \frac{16 \pi}{5e^{2}} B(E2,0^{+} \to 2^{+})
\end{equation}
and
\begin{equation}
 C \approx 0.37
\end{equation}
with $b=15$~fm for an energy of $49.2 \cdot A$~MeV.
{}From Fig.~1a of Ref.~\cite{Bau89}
it can be seen that higher order effects are negligible for the value
of the strength parameter $C$.
This is in agreement with the result found in the coupled channel
calculation of Ref.~\cite{Mot95a}.
\par
In conclusion, we provide a framework to apply
corrections to 1${}^{st}$ order
electromagnetic excitation. It is appropriate for fast collisions.
We constructed operators which take the influence
of intermediate states into account.
This can lead to a great simplification as compared, e.g., to
the coupled channels approach. In this approach, a set of states,
considered to be relevant, has to be chosen with known electromagnetic
matrix-elements. In the present approach, of course, the model
dependence cannot be altogether avoided; it enters when the corresponding
matrix-elements of the operator has to be calculated, or at least
estimated.
\par
As an example we studied the excitation of the $\frac{1}{2}^{-}$
state in ${}^{11}$Be. The importance
of higher order effects for this case of an extremely loosely
bound nucleus was established.
The estimate for the reduction of the cross section can only partly
explain the observed reduction of the B(E1)-value in the GANIL experiment.
However, for a final analysis,
more accurate calculations with improved wave functions,
including E2 and nuclear effects in the excitation,
should be performed.
\par
We wish to thank P.~G.~Hansen and M.~G.~Saint-Laurent for
stimulating discussions.
\par
Note added in revision:
In the meantime we got to know about a coupled channel study of the
${}^{11}$Be Coulomb excitation by C.~A.~Bertulani, L.~F.~Canto,
and M.~S.~Hussein. They get very
similar conclusions as compared to our findings.
\clearpage
\clearpage
\section*{Figure Captions}
{\bf Figure~1:}
Reduction factor $r$ for the excitation probability as a function
of the characteristic parameter $z$ defined in Eq.~\ref{zdef} for the
two model wave functions (I: -----, II: - - -). \\
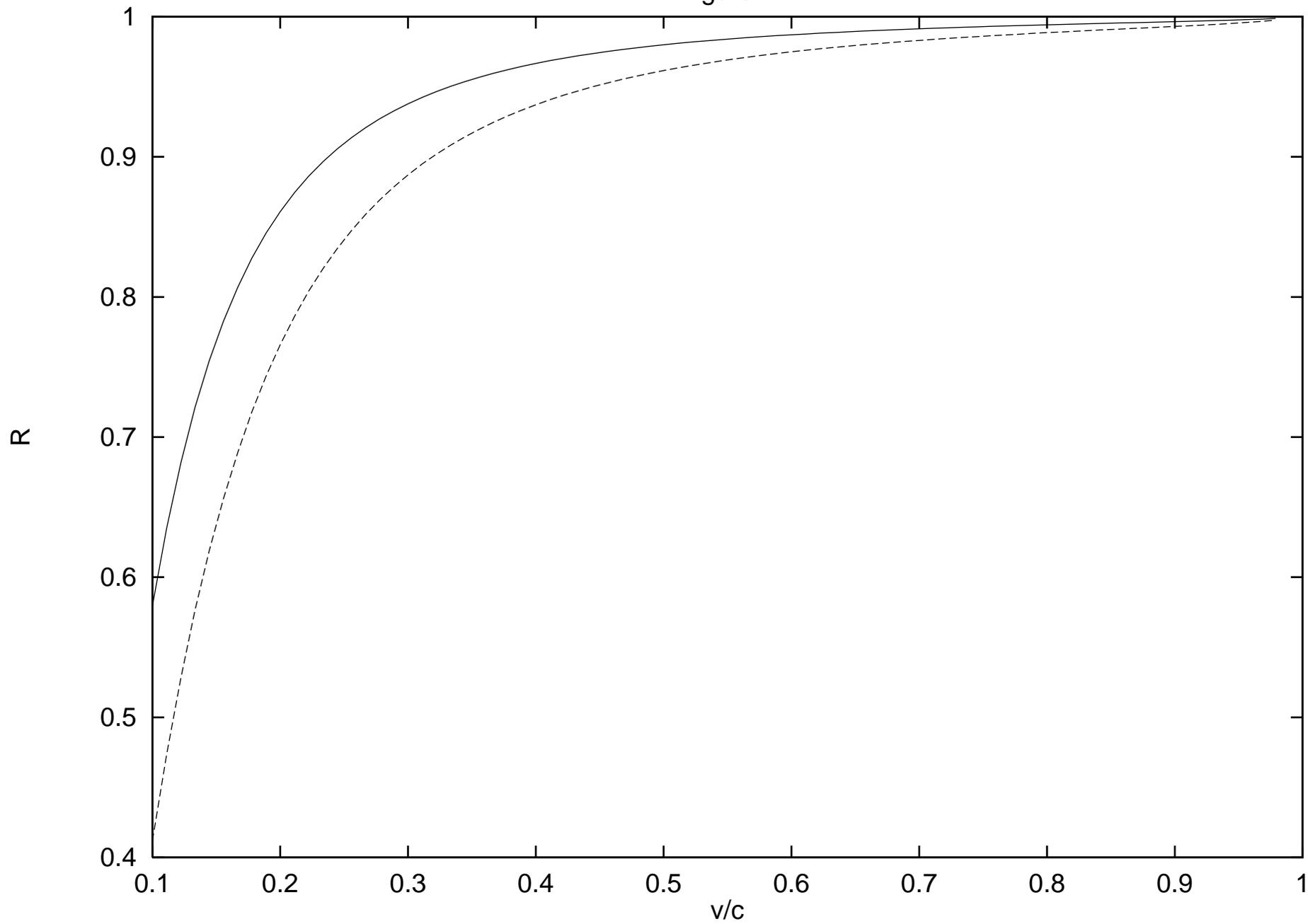
{\bf Figure~2:}
Reduction $R$ of the total cross section for the excitation of
the $\frac{1}{2}^{-}$ first excited state from the $\frac{1}{2}^{+}$
ground state of ${}^{11}$Be as a function of the
projectile velocity $v$ for the two model wave functions
(I: -----, II: - - -).
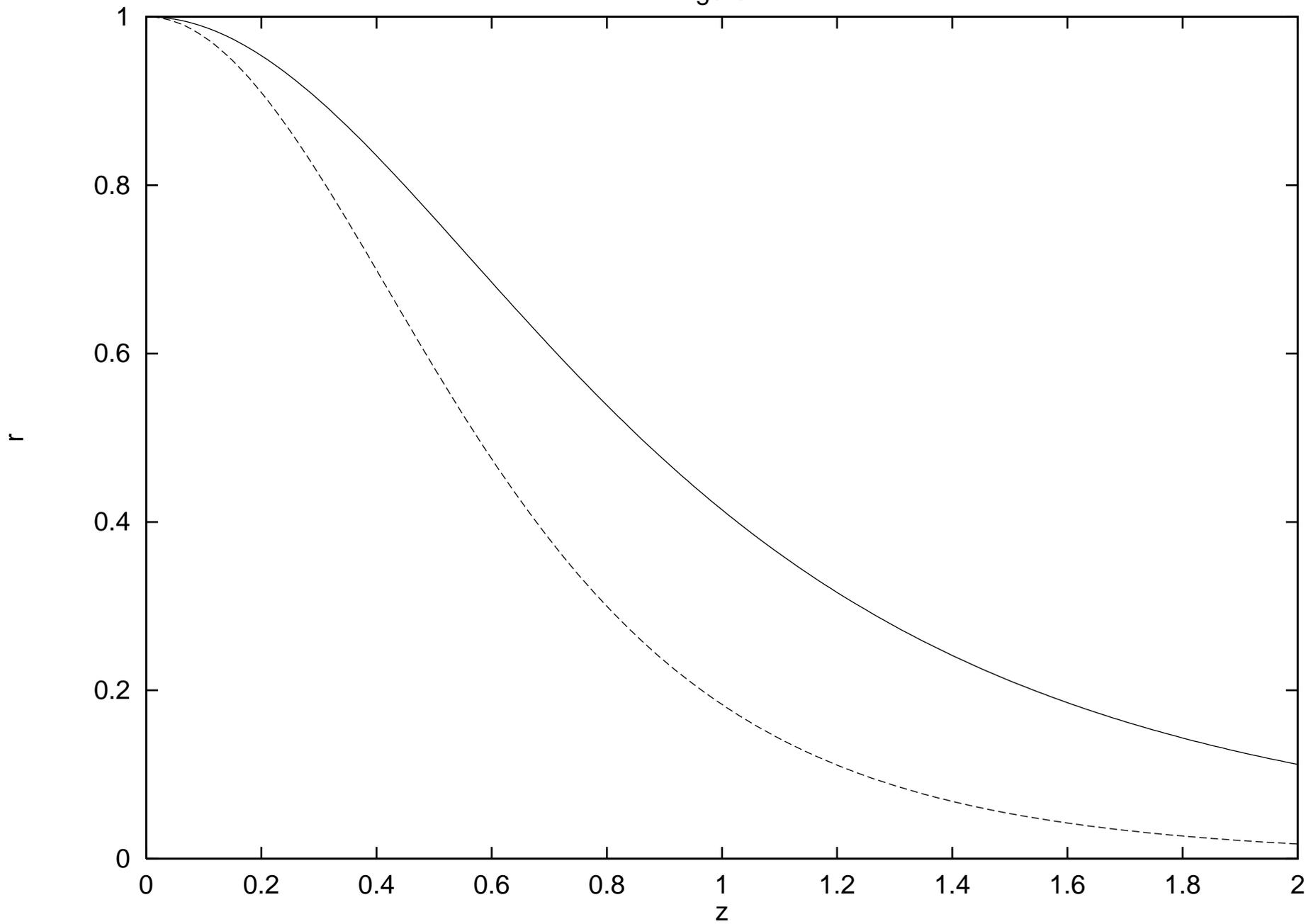
\end{document}